\documentclass[useAMS,usenatbib]{mn2e}
\usepackage{graphicx}

\def\oversim#1#2{\lower0.5pt\vbox{\baselineskip0pt \lineskip-0.5pt
     \ialign{$\mathsurround0pt #1\hfil##\hfil$\crcr#2\crcr\sim\crcr}}}
\def\gsim{\mathrel{\mathpalette\oversim>}}    
\def\lsim{\mathrel{\mathpalette\oversim<}}    
\title[Star and nebula  of  CK Vul]
{The enigma of the oldest
'nova': the central star and nebula of CK Vul }

\author[Hajduk et al.]
  {M.~Hajduk,$^1$
  Albert A.~Zijlstra,$^2$  P.A.M.~van Hoof,$^{3}$ J.~A. Lopez$^4$, 
J.E. Drew$^5$, 
\newauthor 
 A.~Evans,$^6$   S.P.S.~Eyres,$^7$   K. Gesicki,$^1$
R. Greimel,$^{8}$ F.~Kerber,$^9$
\newauthor   S. Kimeswenger$^{10}$
and M.G. Richer$^4$
 \\
        $^1$Centrum Astronomii UMK, ul.Gagarina 11, PL-87-100 Torun, Poland \\
        $^2$University of Manchester,
          School of Physics \&\ Astronomy, P.O. Box 88,
          Manchester M60 1QD, UK\\
	$^3$Royal Observatory of Belgium, Ringlaan 3, Brussels, Belgium\\
        $^4$Instituto de Astronomia, Universidad Nacional Autonoma
            de Mexico, Apdo. Postal 877, 22800  Ensenada, BC, Mexico \\
        $^5$Imperial College of Science, Technology and Medicine, 
         Blackett Laboratory, Prince Consort Road, London, SW7 2BW, UK\\
	$^6$Department of Physics, School of Chemistry and Physics, Keele
	University, Staffordshire ST5 5BG, UK\\
	$^7$Centre for Astrophysics, University of Central Lancashire, Preston
	PRI 2HE, UK\\
        $^{8}$Isaac Newton Group of Telescopes, Apartado de corras 321, 
        E-38700 Santa Cruz de La Palma, Tenerife, Spain\\
	$^9$European Southern Observatory, 
	Karl-Schwarzschild-Strasse 2, D-85748 Garching, Germany\\
	$^{10}$ Institut f\"ur Astro- und Teilchenphysik,
         Universit\"at Insbruck, 
        Technikerstr. 25, 6020 Innsbruck, Austria\\
}

\begin{document}

\date{Accepted Received; in original form }

\pagerange{\pageref{firstpage}--\pageref{lastpage}} \pubyear{2005}

\maketitle

\label{firstpage}

\begin{abstract}
CK Vul is classified as, amongst others, the slowest known nova, a hibernating
nova, or a very late thermal pulse object. Following its eruption in AD 1670,
the star remained visible for 2 years. A 15-arcsec nebula was discovered in
the 1980's, but the star itself has not been detected since the eruption. We
here present radio images which reveal an 0.1-arcsec radio source with a flux
of 1.5\,mJy at 5\,GHz. Deep H$\alpha$ images show a bipolar nebula with a
longest extension of 70 arcsec, with the previously known compact nebula at
its waist.  The emission-line ratios show that the gas is shock-ionized,
at velocities $>100\rm \,km\,s^{-1}$. Dust emission yields an envelope mass
of $\sim 5 \times 10^{-2}\,\rm M_\odot$.  Echelle spectra indicate outflow
velocities up to 360\,km\,s$^{-1}$.  From a comparison of images obtained in
1991 and 2004 we find evidence for expansion of the nebula, consistent with an
origin in the 1670 explosion; the measured expansion is centred on the radio
source. No optical or infrared counterpart is found at the position of the
radio source. The radio emission is interpreted as thermal free-free emission
from gas with $T_{\rm e}\sim 10^4$\,K.  The radio source may be due to a
remnant circumbinary disk, similar to those seen in some binary post-AGB
stars.  We discuss possible classifications of this unique outburst, including
that of a sub-Chandrasekhar mass supernova, a nova eruption on a cool,
low-mass white dwarf, or a thermal pulse induced by accretion from a
circumbinary disk.
\end{abstract}


\section{Introduction}

The enigmatic CK Vul \citep{Shara1982} is currently the oldest catalogued nova
variable.\footnote{Of the eruptive variables, it is predated only by the
well-known Galactic supernovae and by the LBV eruptions of P Cyg in 1600 and
1655} Its classification remains controversial. The star was first observed by
P\`{e}re Dom Anthelme and was discovered independently by Hevelius in
June/July 1670. The observed eruption lasted for two years, during
1670-1672. It was discovered at 3rd magnitude, and faded over 100 days to
below visual limits (magnitude 6).  The following year it was recovered during
rebrightening, reaching magnitude 2.6, 300 days after the first maximum. It
again faded on the same time scale of 100 days to below observational limits.
It showed a faint tertiary maximum 600 days after the first peak, at magnitude
5.5 \citep{Shara1985}.  The star has not been recovered since. Its light curve
is unlike any catalogued novae, raising doubts on its status. Interpretations
include a hibernating nova, or a helium flash on a new white dwarf.

 The historic observations provided coordinates for CK Vul accurate to within
a few arcminutes. Based on the old charts, \citet{Shara1982} were successful
in searching for the remnant. They discovered several nebulosities on their
H$\alpha$+[N\,{\sc ii}] image, the spectra of which present reddened emissions
of [N\,{\sc ii}], H\,{\sc ii}, [O\,{\sc ii}] and [S\,{\sc ii}]. On their
spectra, the flux of [N\,{\sc ii}] $\lambda$ 6584\AA\ is three times more
intense than the H$\alpha$ line, indicating an evolved, and perhaps hydrogen
deficient chemical composition.  No obvious exciting star was found in the
field of CK Vul. A central star candidate was suggested by \citet{Shara1985},
but shown to be misidentified as such by \citet{Naylor1992}; the latter
authors were unable to identify, among the stars visible on their R image, an
alternative central star candidate which could be responsible for the
ionization of the observed nebulosity.

The lack of bright candidates for the central star would constrain the
possible companion of the white dwarf in cataclysmic systems, assuming that CK
Vul is an ordinary nova.  However, the list of features uncommon for classical
novae is extensive.  The light curve during the outburst of  Nova Vul 1670
is unprecedented, showing variations which resemble a set of declines and
subsequent recoveries. The rebrightenings may indicate constant intrinsic
brightness with intermittent dust formation, or a light echo. In the first
instance, the object remained at high luminosity for over a year, with the
fading(s) caused by intermittent extinction. In the second case, a dense
medium needs to be present within a parsec of the source.  The properties of
the shell are quite unusual: a  low outflow velocity of the ejected
material, low density and high ionized mass (although the
distance is very uncertain).  \citet{Shara1985} discuss (and reject) some
alternative hypotheses concerning the nature of CK Vul.  \citet{Harrison1996}
suggests that CK Vul may be a Very Late Thermal Pulse event (VLTP) rather than
a nova: this is based on similarities of the light curve to those seen in
V605 Aql and V4334 Sgr (Sakurai's Object) \citep{Hajduk2005}.
\citet{Evans2002a} have added more arguments for this reborn scenario for CK
Vul.  If the VLTP scenario for CK Vul is correct, it would be only the third
object observed during a VLTP.  \citet{Kato2003} suggested CK Vul to be a
stellar merger.  However, none of the presented scenarios is fully convincing:
the observational data are limited, but the object seems to be unique.

In order to test the VLTP hypothesis we searched for evidence of ionized
material, since these stars reach high temperatures soon after the
eruption and ionize a compact shell.   In contrast, radio emission 
from classical novae fades within a few years after the eruption.
To this end we conducted radio observations and also obtained deep
images, to search for the ejecta, and for evidence of a fossil
planetary nebula. 

\section[]{Observations}

\subsection{Radio emission}

The first radio observations of CK Vul were carried out with the VLA array on
the $4^{\rm th}$ of April 2005. The VLA was in the so-called 'B'
configuration, with a largest baseline of about 10\,km. The observations were
done at a frequency of 5\,GHz. CK Vul was observed for approximately 2 hours,
divided into $\sim$14-min scans, interspersed with observations of the phase
calibrator (1922+155). A second set of observations was obtained on April 9,
2006, with the VLA in the 'A' configuration.  After each $\sim$10 min scan of
CK Vul, at 5\,GHz and 8\,GHz, alternately, the phase calibrator 1922+155 was
observed.

1331+305 was used as the primary flux calibrator for both the 2005 and 2006
observations. Observational parameters are summarized in Table \ref{vla}.

\begin{table*}
\caption[]{\label{vla}Observational parameters for the VLA observations
(naturally weighted images). The radio source is at position (J2000)
 $19^{\rm h}\, 47^{\rm m}\, 
38.074^{\rm s} \, +27^{\rm o}\, 18^\prime\, 45.16^{\prime\prime}$
  }
\begin{flushleft}
\begin{tabular}{lllllllll}
\hline
  Date & Freq. & $t_{\rm int}$ & Beam  & pa & flux  & $\sigma$ & 1922+155 flux &
  1331+305 flux \\ 
       &       &               & FWHM  & deg & mJy & mJy/beam & mJy           & 
  Jy \\ 
\hline \\
 4 Apr 2005 & 5\,GHz & 2\,h   & 1.52''$\times$1.36'' & $-$54 & 1.46 & 0.016 & 682 &
 4.66\\
 9 Apr 2006 & 5\,GHz & 1\,h   & 0.46''$\times$0.38'' & $+$5 & 1.27 & 0.030 & 660 &
 4.66\\
 9 Apr 2006 & 8\,GHz & 1\,h   & 0.26''$\times$0.23'' & $-$1 & 1.53 & 0.023 & 651 &
 7.34\\
\hline
\end{tabular}
\end{flushleft}
\end{table*}

The data were reduced using standard procedures implemented in the AIPS
package, by Fourier transforming the visibility data
followed by the removal of the dirty beam pattern.  The data were convolved
with a synthesized gaussian point spread function (beam) of the same FWHM as
the dirty beam: the beam sizes are listed in Table \ref{vla}. Low spatial
resolution images were also created, in order to search for confusing sources:
only a few such confusing sources were found at 5\,GHz and 8\,GHz. Both
uniformly and naturally weighted images were made, using the CLEAN
algorithm. The RMS noise on the maps of CK Vul followed closely the
theoretical thermal noise.

\begin{figure}
\includegraphics[width=84mm, clip=]{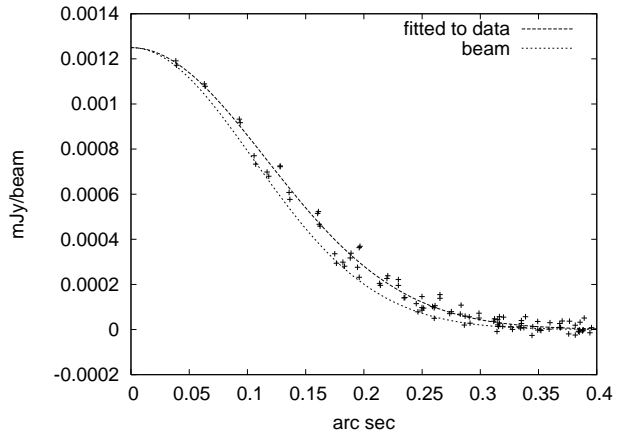}
\caption{\label{profile} The profile of the radio source, based on the
higher resolution 8\,GHz 2006 observations.}
\end{figure}

Our 2005 observation reveals radio emission of $1.46 \pm 0.05$\,mJy (where
the uncertainty is taken as the 3-$\sigma$ noise on the map). The emission is
consistent with a point source: the upper limit of the diameter, deconvolved
from the beam, is 0.5 arcsec. But the higher resolution 2006 observations find
the object to be resolved. The profile of the source, and that of the beam are
shown in Fig \ref{profile}. The FWHM of the beam is $0.26\times0.23$ arcsec,
the convolved source $0.29\times0.26$ arcsec, and the deconvolved FWHM
(gaussian fit) is $0.12\pm0.01\times0.11\pm0.01$ arcsec. The fitting rules out
a point source.

The fluxes are listed in Table \ref{vla}: they were measured both from
gaussian fitting and from summing. The slight difference between the two
epochs may be due to calibration uncertainties or intrinsic variability. The
peak intensity of the 8\,GHz emission (natural weighting) is
$1.25\pm0.02$\,mJy/beam, lower than the integrated flux. This also confirms
that the source is resolved. The source is not circularly polarized, but due
to the faintness of the source the detection limit for polarization is only
about 20 per cent.

\subsection{H$\alpha$ observations}

\subsubsection{INT}

The field of CK Vul was observed on two different occasions at the Isaac
Newton Telescope (INT), located on La Palma.  The first observations were
carried out during a service night on Aug 4, 2004, and the second were obtained
as part of the IPHAS survey, on July 12, 2005.  IPHAS is a CCD H$\alpha$
imaging survey of the Northern Galactic plane \citep{Drew2005}.  This survey
uses 120-sec exposures through the H$\alpha$ filter, 30 sec through the Sloan
$r^\prime$ and 10 sec through Sloan $i^\prime$ (every sky position is covered
twice).  The deeper service observations used three 600-sec H$\alpha$
exposures, supplemented by three 90-sec exposures with the Sloan $r^\prime$
filter.  All observations used the wide field camera, which has a 30 arcmin
field of view, mapped onto a mosaic of 4 CCDs. The pixel size is 0.3 arcsec.

The observations were taken using the IPHAS observing procedures and
were reduced using the IPHAS pipeline: the procedure includes bias
subtraction, flat-fielding, non-linearity correction and astrometric fit. The
emission associated with CK Vul was centred on one of the four CCDs. The
pipeline also obtains photometry of all point sources in the field, stored in
a catalogue.  The 2004 observations were taken under non-photometric
conditions, with intermittent high cloud.

The $r^\prime$ filter is used for the continuum subtraction, but it should be
noted that its broad coverage (5500--7000\AA) also covers H$\alpha$.  The
H$\alpha$ filter has a width of 95\AA, and includes the [N\,{\sc ii}]
lines. An emission line image was obtained from the difference of the
H$\alpha$ and $r^\prime$ images.

The H$\alpha$ and $r^\prime$ observations have different values for the image
quality.  A direct subtraction of the $r^\prime$ image therefore leaves
significant stellar residuals--we note that the field is very crowded. Both
positive and negative residuals are also caused by the fact that stars show
variable $r^\prime - $H$\alpha$ colours due to the stellar H$\alpha$
absorption, and by TiO bands for M stars. This can affect the $r^\prime-\rm
H\alpha$ colour by as much as a magnitude.  To circumvent these problems,
continuum subtraction was done in a different way. We used the DAOPHOT program
to obtain a list of stellar positions for the $r^\prime$ image. This list was
subsequently used to find the counterparts of these sources in the H$\alpha$
image. A point-spread function fitting was done to obtain the magnitude (in
H$\alpha$), and to subtract these stellar sources. The same fitting and
subtraction procedure was also carried out in the $r^\prime$ image. Finally,
the residual $r^\prime$-band image was subtracted from the residual H$\alpha$
image. This procedure greatly reduces the stellar residuals but may subtract
true H$\alpha$ emission stars. We checked for the presence of such stars in
the area surrounding CK Vul but did not find any.

The astrometry of the image was re-calibrated using the positions of the stars
in the Second US Naval Observatory CCD Astrograph Catalog 
 \citep[UCAC2][]{Zacharias2004}. The accuracy of the astromery,
based on the RMS of the fit, is 0.1 arcsec.

Below we refer to the final image as 'the H$\alpha$ image' (adopting the
name of the filter). However, the reader should be aware that the
[N\,{\sc ii}] lines may contribute or even dominate the line emission
within the filter pass band.

\subsubsection{WHT}

H$\alpha$ images have previously been presented by \citet{Naylor1992}. The
original data were kindly provided to us and re-reduced using IRAF.  The
observations were carried out at the William Herschel Telescope, on La Palma,
on August 10, 1991. They made use of a small CCD at the auxiliary Cassegrain
port; the scale is 0.1 arcsec/pixel. The observations in H$\alpha$ were
de-biased and flat-field corrected. The image was calibrated with respect to
the positions of the stars on the INT image: the RMS of the relative fit is
0.05 arcsec.

\subsection{Long-slit spectrometry}

The long-slit observations were obtained with the Man\-ches\-ter echelle
spectrometer \citep[MES:][]{Meaburn2003} combined with the f/7.9 focus of the
2.1 m San Pedro M\'artir, UNAM telescope, on 2006 July 12. This echelle
spectrometer has no cross-disperser. For the present observations a filter of
90\AA\ bandwidth was used to isolate the 87$\rm ^{th}$ order containing the
H$\alpha$ and [N\,{\sc ii}] emission lines. A  $1024 \times 1024$
SITE CCD with 24$\mu$m pixels was used. Seeing conditions were
variable, with a mean of 2 arcsec during the observations.

Four slit positions across CK Vul were obtained. The first two positions
employed an on-chip binning factor $2 \times 2$ and a slit width of 150$\mu$m,
equivalent to 11\,km\,s$^{-1}$ and 0.6 arcsec increments along the
slit. However due to the faintness of the emission and the relatively poor
seeing, the third and fourth positions were obtained with a $4 \times 4$
binning and a 300$\mu$m wide slit corresponding to 20\,km\,s$^{-1}$
resolution and 1.2 arcsec increments along the slit for better
sensitivity. The slit orientations of the latter are shown in Fig. \ref{slits}
and the spectra in Fig. \ref{mosaic}. The slits were oriented in all cases
N--S and the integration times were 1800 seconds. The spectra were reduced and
calibrated in wavelength against a Th--Ar lamp using standard procedures.

\subsection{Other data}

An earlier radio continuum observation of CK Vul at 20 cm was performed by
\citet{Bode1987}, on January 28, 1984. They found an upper limit of
1.52 mJy, with a spatial resolution of 14 arcsec. Due to the low resolution,
this limit includes both the compact source and the extended nebulosity.

Proposed R-band and JHK counterparts are summarised in \citet{Evans2002a}. We
consider these to be uncertain, referring to unrelated (field) stars. 2MASS
finds all these to have the colours of normal stars.  Observations in other
wavelength regimes include flux measurements at 850 and 450$\mu$m and flux
densities derived from re-analysis of the IRAS data \citep{Evans2002a}. Upper
limits were found for MSX (8.3 $\mu$m) and IRAS 100 $\mu$m.

 GLIMPSE/Spitzer images of the field confirm that all 2MASS sources show
stellar Rayleigh-Jeans colours. This includes the star inside the cavity south
of the radio source, which was studied by \citet{Harrison1996}.

\section{Results of the observations}

\subsection{Optical lobes}

\begin{figure}
\includegraphics[width=84mm, clip=]{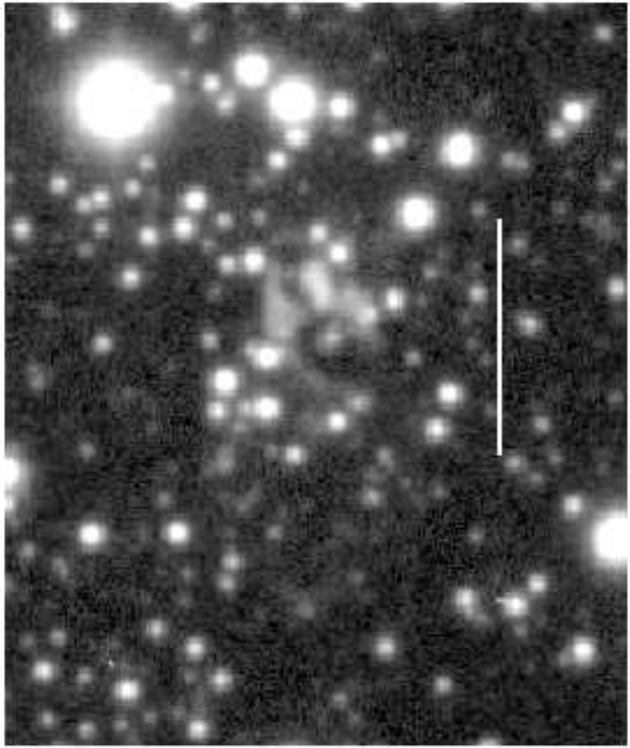}
\caption{\label{ha} The INT H$\alpha$ image of CK Vul, before
continuum subtraction. A logarithmic intensity scale is used to bring out
the faint emission. The 
vertical bar is 30 arsec in length; North is at
the top and East is left.}
\end{figure}

The H$\alpha$ image, derived from the Aug. 2004 (long) integrations, is shown
in Fig. \ref{ha}.  It shows the dense star field, and the compact structure
first found by \citet{Shara1985}. The nebulosity appears as a partial ring,
extending 15-arcsec across E-W. The star at the centre of the ring,
originally suspected to be CK Vul, is now thought to be a field star. Faint
nebular emission is lost in the dense star field.

\begin{figure}
\includegraphics[width=80mm, clip=]{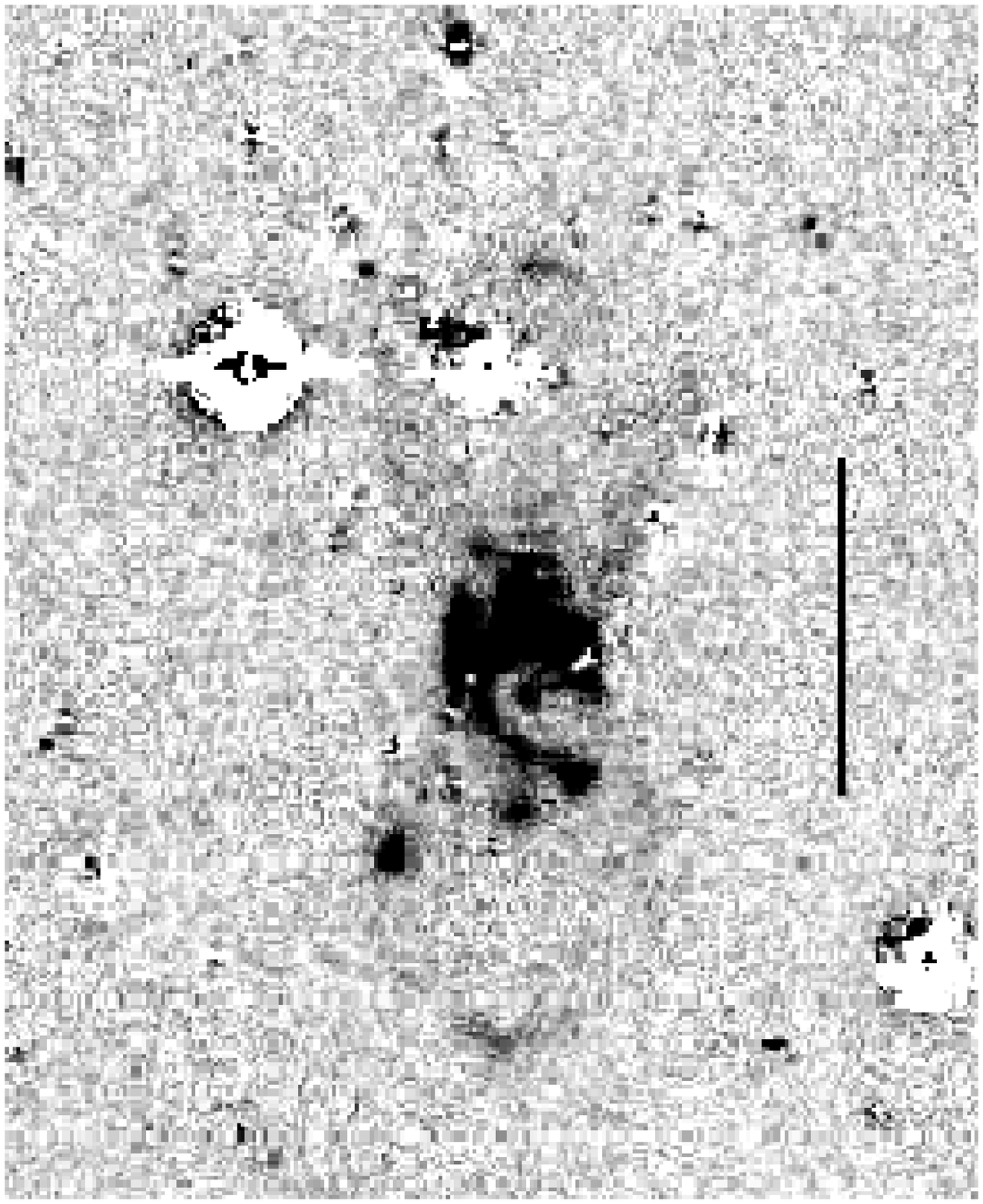}
\includegraphics[width=80mm, clip=]{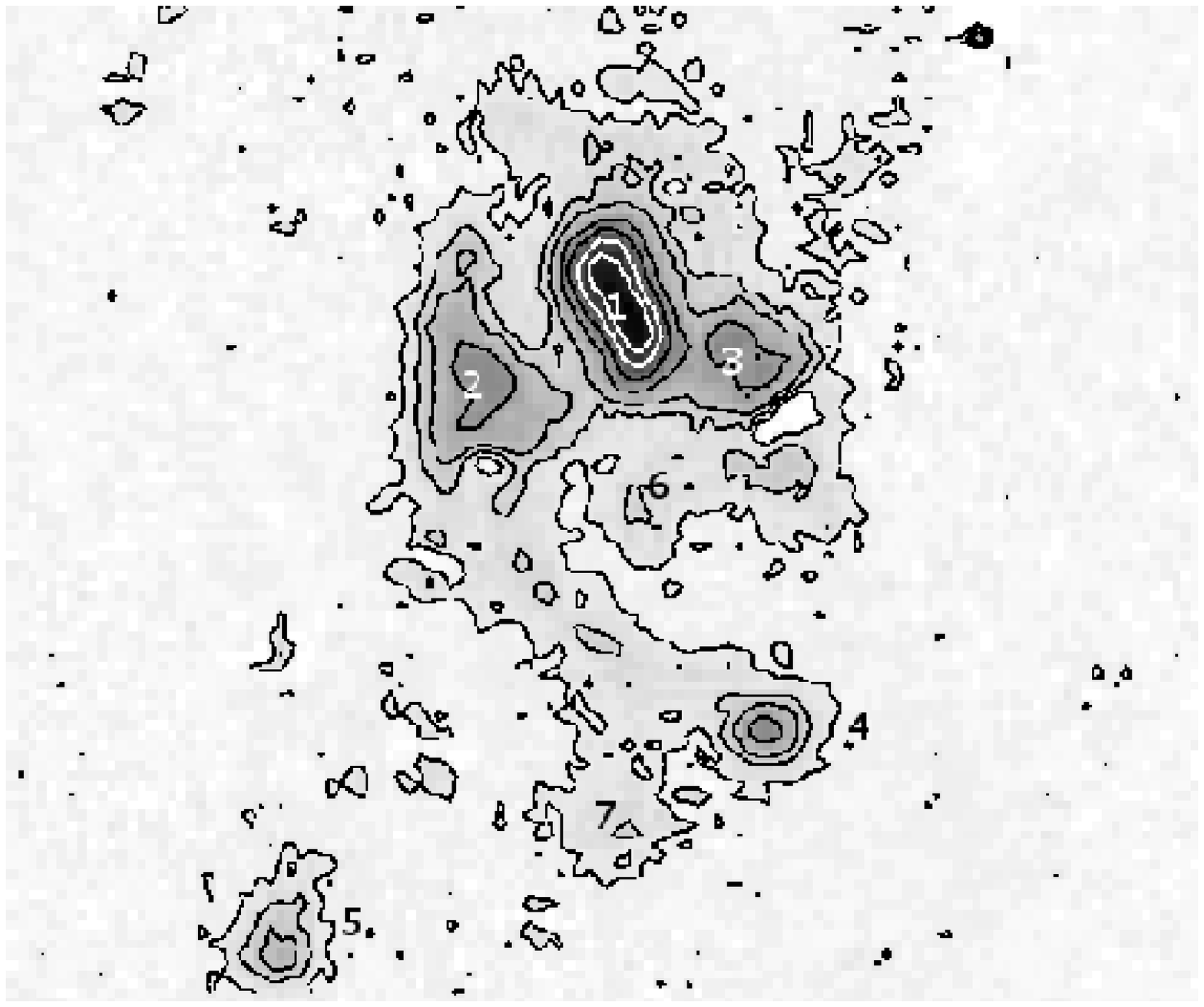}
\caption{\label{iphas_nebula} The continuum-subtracted INT H$\alpha$ image of
CK Vul.  North is top, east is left. A (Top): Image encompassing all detected
line emission.  The contrast is set to bring out the faint bipolar lobes. The
vertical bar is 30 arsec in length.  B (Bottom): The central region,
corresponding to the emission detected
by \citet{Shara1985}.  The linear contours
and the grey scale refer to the same image.  The peaks on
either side of the apparent 'funnel' are 5 arcsec apart. The numbers 1 to 5
correspond to the knots detected by \citet{Shara1985}. Numbers 6 \&\ 7 have 
been added in this paper; see text for discussion.}
\end{figure}

The corresponding continuum-subtracted optical image is presented in
Fig. \ref{iphas_nebula}. The image shows that the previously known nebulosity
is located at the waist of a much larger, bipolar structure. The bipolar lobes
are edge-brightened: the filled emission is just visible in the northern lobe
but is below the detection limit in the southern lobe. A number of small
condensations are visible, two of which delineate the tips of the lobes. A
bright condensation is seen on the SE lobe. The angular distance between the
tips of the lobes is 70 arcsec. The filter transmission includes both H$\alpha$
and [N\,{\sc ii}], and either line may contribute in different areas of the
structures.

The central region is shown in more detail in Fig. \ref{iphas_nebula}B.  A
funnel-like structure is seen, with one side much brighter than the other.
(But whether the structure is in fact a cavity is not known.)  Four other
knots are visible: each of these components are numbered on the image as in
\citet{Shara1985}. An arc connects knots 2 and 4, but it is not part of a
closed ring. The three-dimensional structure cannot easily be deduced from the
image. In addition to the numbering of \citet{Shara1985}, we number (6) the
central diffuse source and (7) the more diffuse emission between knots 4 and
5.

We note that a stellar source on the western ring was subtracted in the
continuum removal process. This left the blank residual below knot 3. Some
compact H$\alpha$ emission was likely subtracted in this process, as 
suggested by the residuals. This problem occurs when a star is much
brighter than coincident H$\alpha$ emission, but dos not affect other
locations within the nebulosity.  Fig. \ref{radio}A
shows the inner region of $r^\prime$-band INT image, where this star
is visible to the West of the radio source.

\subsection{Radio source}

\begin{figure}
\includegraphics[width=84mm, clip=]{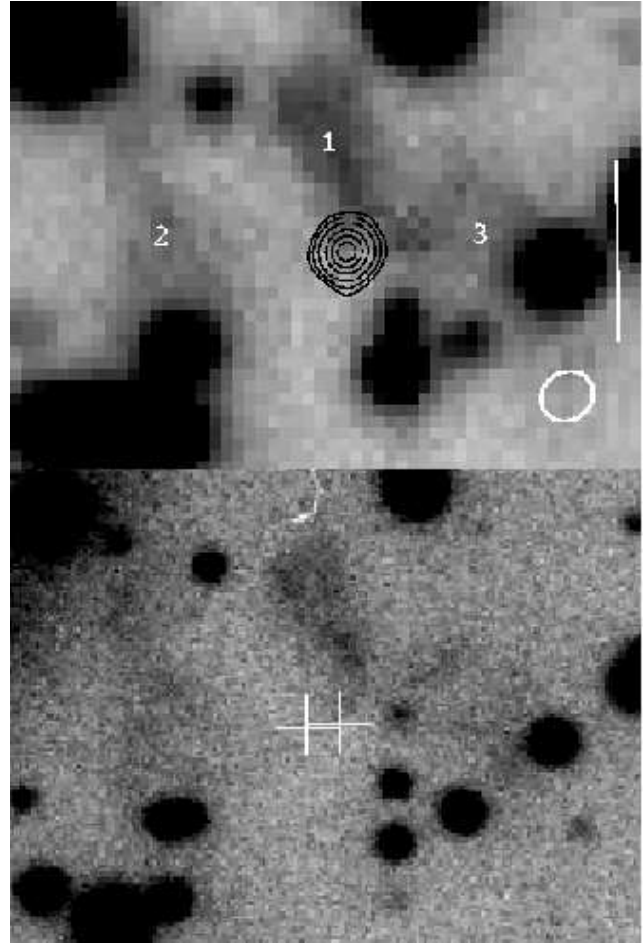}
\caption{\label{radio} The radio source superposed on the optical
images. North is top, east is left. The position of the radio source is
$19^{\rm h}\, 47^{\rm m}\, 38.074^{\rm s} \, +27^{\rm o}\, 18^\prime\,
45.16^{\prime\prime}$. The knots are indicated by their numbers. A (top):
superposed on the $r^\prime$-band INT image. The image is 21.5 arcsec across
horizontally.  Radio contours are at 0.075, 0.15,0.3,0.6,0.9,1.2 mJy/beam,
from the 2005 VLA observations (natural weighting). The oval indicates the
size of the beam; the black bar is 5 arcsec long. B
(bottom): superposed on the WHT H$\alpha$ image. The right-most plus sign
indicates the position of the radio source. the left-most plus sign gives
the geometric centre of the bipolar lobes.
}
\end{figure}

 The radio emission is shown in Fig. \ref{radio}A superposed on the INT
$r^\prime$-band image.  The brightest emission-line regions are faintly
visible in the $r^\prime$-band, because the wavelengths of the emission lines
fall within the filter transmission curve.  The radio source is located
between knots 1 and 6 (Fig. \ref{iphas_nebula}).  Three stars located in the
region below the funnel could be candidates for the central star, but they do
not coincide with the compact radio source.  The radio source also does not
coincide with any of the H$\alpha$+[N\,{\sc ii}] bright knots.

An accidental superposition of a background radio source is improbable, due
to the excellent coincidence of the position of the radio emission with the
nebula. Moreover, only one other source with comparable flux to the one
referred to CK Vul was found in the full primary beam field (FWHM) of 9.0
arcmin.

The radio source is close to the brightest part of the elongated knot '1'.
Fig. \ref{radio}B shows the WHT image, which has better spatial resolution.
A faint elongation or separate source is indicated just below knot '1'; on the
INT image this merges with the knot because of the worse seeing.  The faint
elongation is within 0.5 arcsec of the radio source (right-most plus sign in
Fig. \ref{radio}B). It has a very faint $r^\prime$-band counterpart, confirming
it is an emission line region. To classify this extension as a separate source
would require either higher angular resolution, or kinematic data. Given the
extent of the emission line regions around the radio source, the likelihood of
a chance superposition is rather high.  However, it is our only --but
speculative-- candidate for an optical counterpart.

 The radio source is situated in a cavity close to the brightest part of the
nebulosity.  Nebulosity located further from the radio source shows a tendency
of decreasing brightness with distance from the radio source.  This indicates
that the radio source is associated with the current energy source of the
nebula, and reveals the position, if not the identification, of the central
object.

The radio source is to a very good (1-arcsec) accuracy located at the centre
of symmetry of the tips of the bipolar nebula. This point is indicated by the
left-most plus sign in Fig. \ref{radio}B.  This also supports the
identification of the radio source as the central object.

The INT data show that at the position of the radio source, there is no
optical counterpart down to a conservative limit of $r^\prime > 23$. The
i-band IPHAS image is less deep, and gives a limit of $i^\prime>20$.

\subsection{Expansion}

\begin{figure}
\includegraphics[width=84mm]{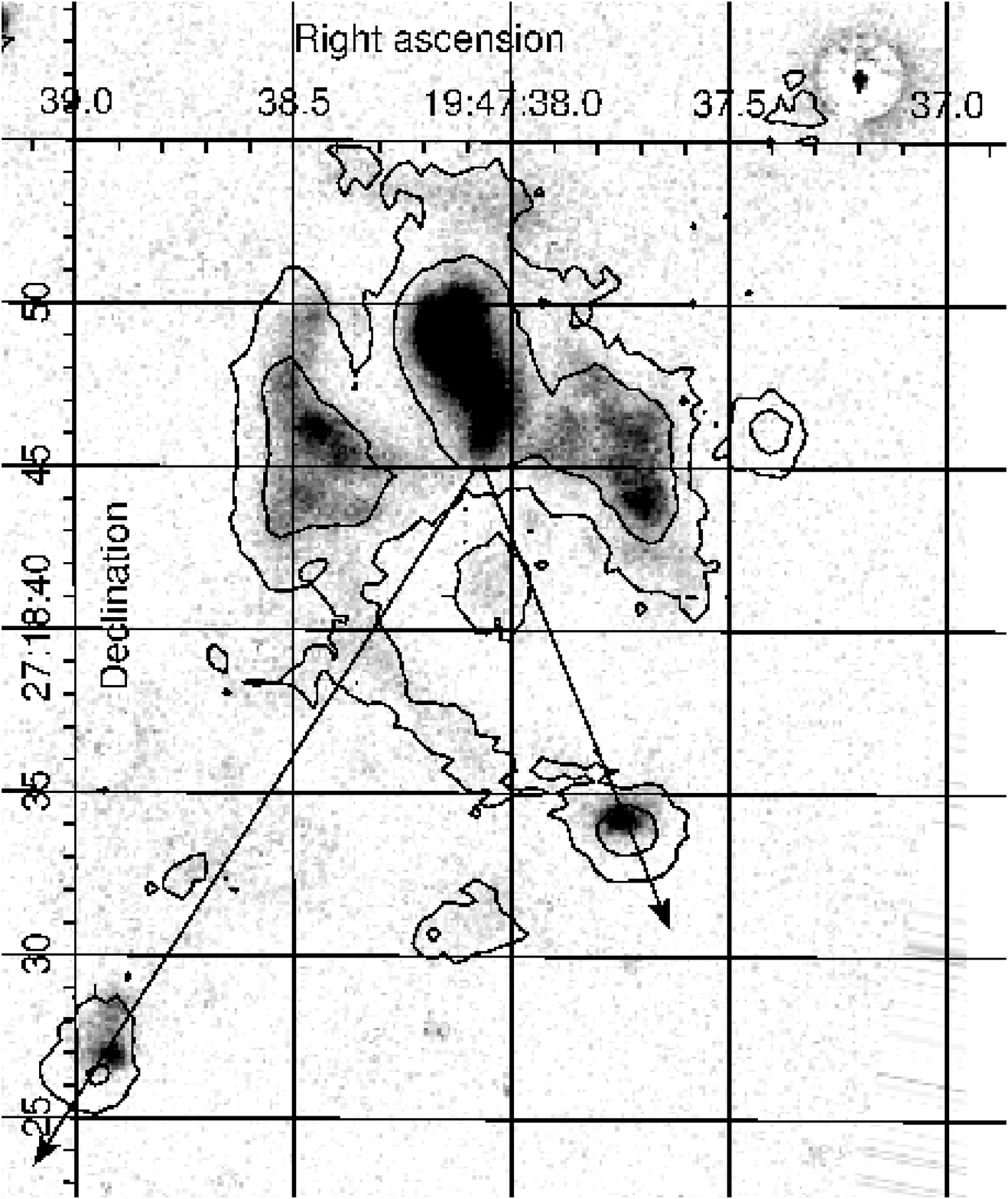}
\caption{\label{ckvul_movingblobs.ps} The WHT H$\alpha$ from 1991 (grey scale)
together with the 2004 INT H$\alpha$ image (contours at 3 and 10
$\sigma$). Two compact components show significant proper motion. The arrows
link the radio sources to these components: the
direction is consistent with the observed proper motion. }
\end{figure}

The WHT image compared to the INT image allows us to search for temporal
variations in the nebula. The WHT data were taken in 1991 and the INT
data in 2004. The 13-year separation represents 4 per cent of the
elapsed time since the AD 1670 eruption.

In Fig. \ref{ckvul_movingblobs.ps}, the grey scale shows the WHT image, and
the INT H$\alpha$ (including continuum) image is overlaid as contours. The WHT
image does not cover the full region of the nebula (which was not known at the
time).  The images are aligned on the stellar positions.  The relative (rms)
accuracy is 0.05 arcsec.

Several emission-line nebulosities have changed position, in particular knots
4 and 5. (The object to the West of knot 3 has moved
north: this is interpreted as a high proper motion star.)  The change in
position is away from the central radio source, as indicated by the
arrows in Fig.  \ref{ckvul_movingblobs.ps}.  The positions of knots 4 and
5 are given in Table \ref{shift}.  The indicated relative motion from 1991 to
2004 is ($\mu _{\alpha},\mu _{\delta}) = (-0.16, -0.46)$ arcsec, and
($\mu _{\alpha},\mu _{\delta}) = (0.36,-0.67)$ arcsec, for knot 4 and
5, respectively.

The original positions of the two components are obtained by extrapolation
back to 1670.  The extrapolated position of knot 4 is coincident with the
position of the radio source, assumed to represent the central star.  The
agreement is worse for knot 5, possibly due to its extended, irregular
structure, which is better resolved in the WHT image than in the INT image
(Table \ref{shift}). Residual image distortions may also be present in the WHT
images.

\begin{table}
\caption[]{\label{shift}Centroid positions (J2000) of the two components
indicated in Fig. \ref{ckvul_movingblobs.ps}. The accuracy of the
determination of centroids is typically about 0.02 arcsec, but for the
component 5 at 2004 is a few times worse. }
\begin{flushleft}
\begin{tabular}{lllllll}
\hline
 Component & & {1991 position} & 
 {2004 position} \\
\hline \\
 4  & $\alpha$ & $19^{\rm h}\, 47^{\rm m}\, 37.753^{\rm s}$ & 
     $19^{\rm h}\, 47^{\rm m}\, 37.741^{\rm s}$  \\
    & $\delta$ & $+27^{\rm o}\, 18^\prime\, 34.31^{\prime\prime}$ & 
     $+27^{\rm o}\, 18^\prime\, 33.85^{\prime\prime}$ \\
 5 &  $\alpha$ & $19^{\rm h}\, 47^{\rm m}\,  38.931^{\rm s}$ & 
     $19^{\rm h}\, 47^{\rm m}\, 38.958^{\rm s}$ \\
    & $\delta$ & $+27^{\rm o}\, 18^\prime\,  27.18^{\prime\prime}$ & 
     $+27^{\rm o}\, 18^\prime\, 26.51^{\prime\prime}$ \\
\hline
\end{tabular}
\end{flushleft}
\end{table}

The observed motion of these two components, located outside the inner
region of the nebula, implies that the observed lobes are indeed the remnants
of the 1670 event. It also confirms our interpretation of the radio source as
the eruptive object. 

To improve the sensitivity to expansion, we subtracted the IPHAS image from
the WHT observations. The subtracted image showed residuals at the position of
the two outer, moving knots, the central brightest nebulosities, the faint arc
and other nebular structures, present on the H$\alpha$ image. Applying a
constant scaling factor prior to the subtraction, to reduce the image scale of
the IPHAS image, greatly reduced these residuals.

\begin{figure}
\includegraphics[width=84mm]{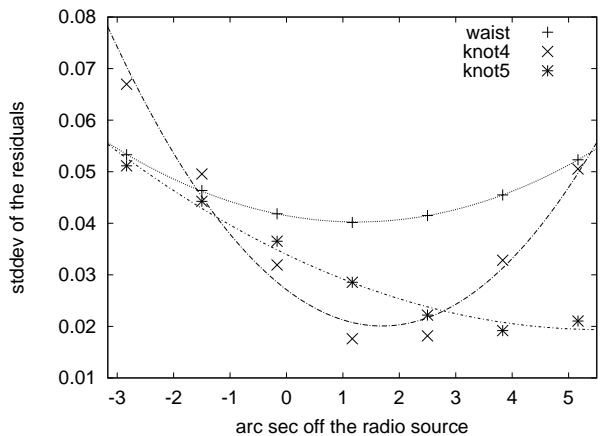}
\caption{\label{stdevy.eps} The residuals of the scaled IPHAS image and 
the WHT  H$\alpha$ images, versus the position of the 
assumed expansion centre. The position of radio source is at 0.
}
\end{figure}

Figure \ref{stdevy.eps} shows the residuals of the two knots, and the
brightest part of the inner nebula, for different positions of the centre of
the expansion. The best fit was obtained for the assumed expansion centre
lying within 2 arcsec of the radio emission. The constant scaling factor
assumes that the material seen in the field  of CK Vul is coming
from the 1670 eruption event and is flowing ballistically.

Unfortunately, the tips of the lobes of the outer nebula are outside of the
WHT frame. However, the tip of the southern lobe is just visible on the plate
published in \citet{Shara1985} (their Fig. 8). Comparison shows that this tip
has moved by roughly 2.5 arcsec in the intervening 20 years. Within the
uncertainties, this proper motion is consistent with a 1670 ejection event.

\subsubsection{Brightness}

\begin{table}
\caption[]{\label{flux}Line fluxes for the different knots,
as measured from the INT image. For knot 3, the flux includes
a non-subtracted star. The values for knots 6 and 7 are also uncertain.
The uncertainty in the flux conversion factor is estimated at 10 per cent.
The [N\,{\sc ii}] 6584/H$\alpha$ ratios are indicative only.
}
\begin{flushleft}
\begin{tabular}{llll}
\hline
Knot & Flux  & $v_{\rm hel}$ & [N\,{\sc ii}]/H$\alpha$ \\
   & [$10^{-15}\,\rm erg\,cm^{-2}\,s^{-1}$] & [km\,s$^{-1}$] \\
\hline
1    &    43  			\\
2    &    28     		& $-80$ & 1.5 \\
3    &    \llap{$\le$}25	& +0 	& 1 \\ 
4    &    4.7    		& $+60$ & $>10$ \\
5    &    3.6   			\\
6    &    1.7    			\\
7    &    2.6    			\\
southern tip &  		& $-175$  & $>2$ \\
\hline
\end{tabular}
\end{flushleft}
\end{table}

We measured fluxes for the different knots as follows: The $r^\prime$
magnitudes of 12 field stars were obtained from the IPHAS catalogues. 
The magnitudes were converted to Bessell filters on Landolt standard
stars using the colour transformations
$$ r^\prime-R_{\rm Lan} = +0.275(R-I)_{\rm Lan}+0.008 $$
$$ (r^\prime-i^\prime) = 1.052(R-I)_{\rm Lan}+0.004 $$
The
magnitudes were converted into flux units per \AA\ using the
\citet{Bessell1998} coefficients for R. The integrated in-band flux was
obtained by multiplying by the equivalent width of the filter (80\AA); this
was divided by the data count (ADU) for each star to get a flux per data
count. We find a conversion value of $(94 \pm 6)\times 10^{-20}\,\rm
erg\,cm^{-2}\,s^{-1}$ per data count. This factor was applied to the data
counts for each of the knots in the continuum-subtracted image.

The results are listed in Table \ref{flux}.  The line fluxes contain
contributions from H$\alpha$, and from the two [N\,{\sc ii}] lines: the filter
transmission is within 5\%\ of the peak transmission for each of the
lines. Based on the spectroscopic data, H$\alpha$ accounts for typically half
to a quarter of the measured line flux, but for knot 4 we do not detect
H$\alpha$.

The changes between the two INT images, measured in the same way, indicate a
0.08\,mag (approx. 10\%) change in the brightness of the knots, but this
varies between knots by more than 10\%. The standard deviation for the
background stars alone is 0.05\,mag, so that we do not see conclusive evidence
for brightness evolution.

The brightest knots are those closest to the radio source.  This suggest that
their excitation arises from a central source.

\subsection{Spectroscopy}

Our echelle spectra show [N\,{\sc ii}] 6584/H$\alpha$ ratios between 1 and
5. Individual values are listed in Table \ref{flux}. Because of the low S/N,
these need confirmation.  Knot 4 has no H$\alpha$ detected and could have a
very high ratio.

These ratios can be compared to other values reported in the literature.  
\citet{Shara1985} analyze spectra of the three brightest nebulosities,
numbered 1--3. They find consistent line intensities, with remarkably strong
[S\,{\sc ii}] emission comparable to H$\alpha$, and a ratio [N{\sc
ii}]6584/H$\alpha \approx 3$. The [S\,{\sc ii}] 6716/6730 line ratio is
approximately two. [O\,{\sc iii}] is also seen, several times
brighter than H$\beta$, but weakened by the extinction.

A spectrum of a faint knot is reported by \citet{Naylor1992}.  The knot has
[N\,{\sc ii}] 6584/H$\alpha \approx 1$, and a [S\,{\sc ii}] 6716/6730 ratio of
unity.  It is not clear which knot this represents: their slit orientation at
position angle 45 degrees covered knot 4 and 6. Our much higher line ratio for
knot 4 suggests Naylor et al. observed knot 6.

A spectrum taken with a 4-arcsec slit oriented perpendicular to this direction
is reported by \citet{Cohen1985}. Their spectrum shows  intermediate values
for both the N and S line ratios compared to the previous two papers.

\subsection{Kinematics}

\begin{figure}
\includegraphics[width=84mm, clip=]{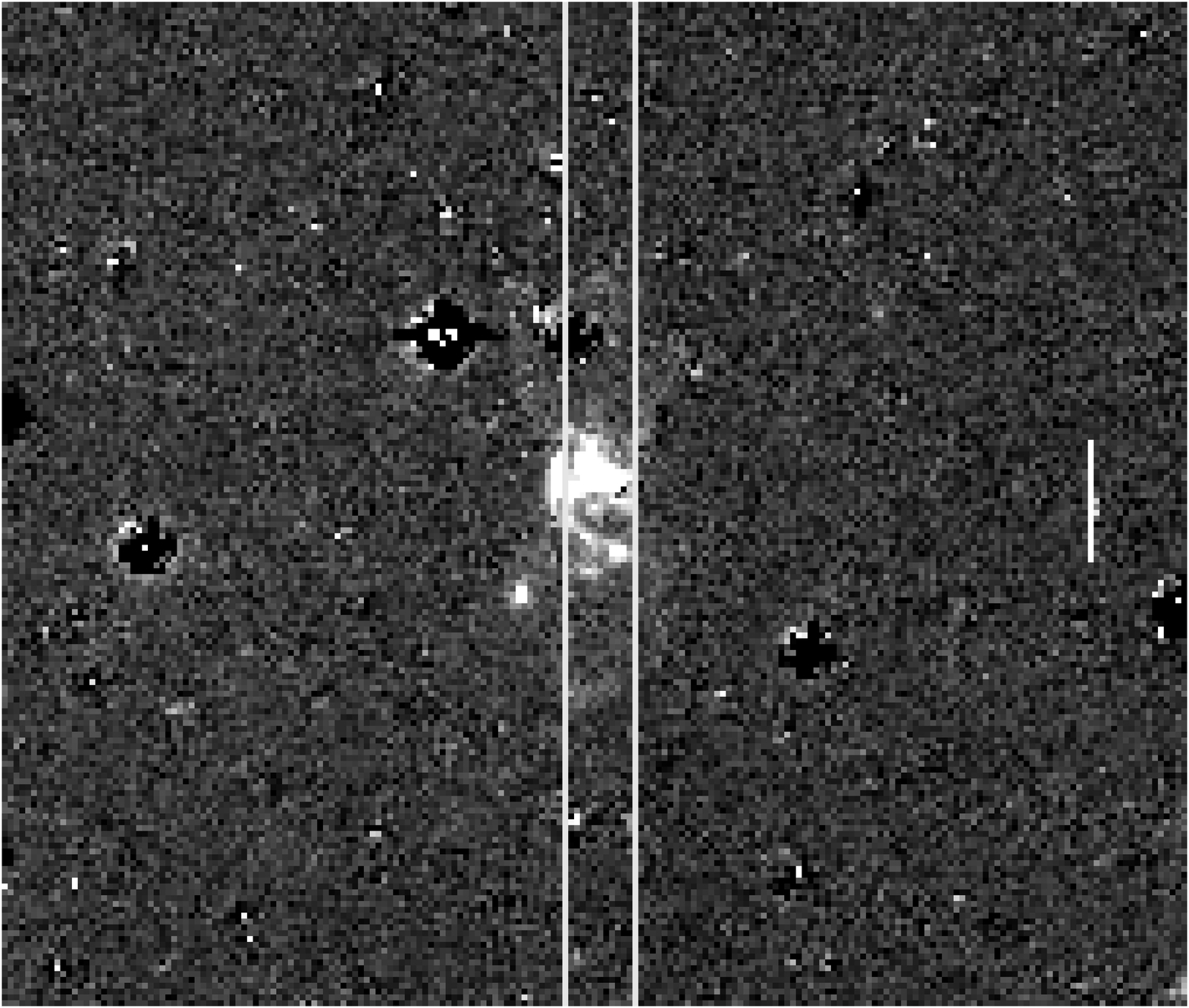}
\caption{\label{slits} Positions of the two long slit echelle spectra
discussed in the text. The scale bar is 20 arcsec long.
 }
\end{figure}

\begin{figure}
\includegraphics[width=84mm, bb= 75 150 780 715]{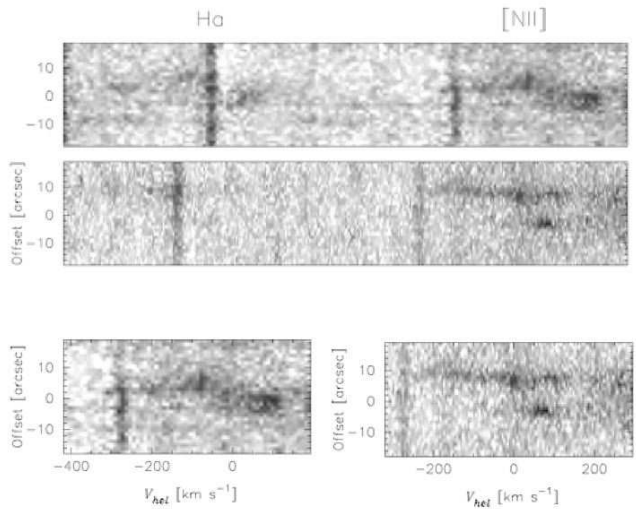}
\caption{\label{mosaic} Two long slit echelle spectra. The upper frames show
the full order, covering H$\alpha$ and [N\,{\sc ii}] 6584\AA. The two lower 
frames show only the  [N\,{\sc ii}] line. Lower left frame:  knot 2 (upper) 
and the southern arc (lower emission). Lower right frame: knot 3 (upper) and 
knot 4 (lower emission).
 }
\end{figure}

The two long-slit spectra with better signal to noise are shown in
Fig. \ref{mosaic}; only signal from the relatively bright knots in the core
was recorded. The top and middle panels in Fig. \ref{mosaic} show both the
H$\alpha$ and [N\,{\sc ii}] 6584\AA\ emission lines for each slit position. 
The locations of the slits are shown in Fig. \ref{slits}. The
first slit (top panel) intersects knot 2 and a section of the southern arc,
whereas the second slit (middle panel) located to the West of the previous
one, intersects knot 3 and the edge of knot 4.  The lower panel shows the
corresponding P--V arrays for the [N\,{\sc ii}] 6584\AA\ line with heliocentric
velocity scales.

The spectra show that the [N\,{\sc ii}] 6584\AA\ line is consistently stronger
than H$\alpha$. Although H$\alpha$ is barely detected, the S/N is enough to
show that its line profile follows the same wide structure of [N\,{\sc ii}] 
6584\AA. The line profiles are peculiar in the sense that the emission seems to
arise only from localised regions and with a wide velocity range, of the order
of $350\rm \,km\,s^{-1}$, somehow akin to those observed in supernovae blasts
\citep[e.g.][]{Riesgo2005}.

The brightest region of each line profile yields the following heliocentric
velocities: for knot 2 we find an approximate velocity of $-80$ km s$^{-1}$;
for knot 3, 0 km s$^{-1}$; for knot 4, $+60$ km s$^{-1}$ and for the southern
arc, $+70$ km s$^{-1}$. From these line profiles the systemic velocity cannot
be determined with confidence. However, since the observed heliocentric
velocities are heavily blue-shifted for knots 2 \& 3, it is reasonable to
assume that there we are detecting material that is mainly coming towards us
whereas the opposite is observed in knot 4 and the southern arc. From this
kinematic behaviour and the morphology of the region (see
Fig. \ref{iphas_nebula}) a possible interpretation is that of a rapidly
expanding (and disrupted) torus or equatorial structure, centred at roughly
$-50$ km s$^{-1}$ and where knots 2 \& 3 are part of the foreground
(approaching) equatorial structure and knot 4 and the southern arc are part of
the background (receding) part of the torus.

The tip of the southern lobe was covered by two of the slit settings. There is
a faint trace in both spectra of a velocity feature at the location of the tip
with a velocity of about $-175\rm\, km\, s^{-1}$. The sign and magnitude of
this velocity is consistent with the bipolar lobes being oriented
perpendicular to the equatorial structure discussed above and with the
northern lobe moving away and the southern one coming towards us.

The kinematics and structure of the line profiles discussed here indicate that
shocks are a main, if not the only, contributor to the excitation of these
emission lines, in agreements with the arguments of the preceding section.

\section{Physical parameters}

\subsection{Extinction distance}

The spectra of \citet{Shara1985} show a strong Balmer decrement, corresponding
to $E(B-V) = 0.82\pm0.23$, or $A(V)=2.5)$. Based on the weighted mean of this
and the extinction towards a distant field star ( $E(B-V) = 0.7\pm0.1$), they
adopt $A(V)=2.2\pm0.3$.

\citet{Shara1985} locate the nebula beyond an extinction layer,
located at $550\pm 150$ pc distance.  \citet{Weight1993} deduce an upper limit
of 2\,kpc: they detect  two CO clumps, with kinematic distances of
500\,pc and 2\,kpc. The two clumps together would result in $E(\rm B-V)=1.25$,
suggesting that CK Vul must lie in front of the more distant clump. 

We have assessed the IPHAS photometry of point sources in the vicinity of CK
Vul to constrain the distance-extinction relation in this field.  Within a
field of 4.5'$\times$4.5', we find a few essentially unreddened main sequence
stars, and then a dearth of stars to $A_V \sim 1.2$ ($E(B-V) \simeq
0.4$). This jump in extinction may represent the dust layer at 550\,pc
described by \citet{Shara1985}.  We have cross-identified a group of 10 early
A stars (presumed to be dwarfs) for which there are 2MASS $JHK$ measurements
as well as IPHAS $r^\prime$, $i^\prime$ and H$\alpha$ data, and used them to
obtain a sample of reddenings and distances consistent with their likely
spectral types.  These early A stars range in brightness between $r^\prime
\sim 14$ and $r^\prime \sim 18$, and are consistent with extinctions in the
range $0.9 \pm 0.2 < A_V <5.4 \pm 0.5$.  The deduced distances fall in the
range 2--4\,kpc.  The reddening does not increase in any organised manner with
distance, and the extinction layer(s) appear to be patchy. 

We will here use a distance of 600\,pc. The extinctions indicate an upper limit
for the distance of CK Vul of 4\,kpc.

\subsection{Velocities}

The radial velocity range is about 350\,km\,s$^{-1}$. The highest blue-shifted
velocities are seen just to the north of knot 2 and may represent the inner
edge of the bipolar lobe, oriented close to the line of sight along the edge
of the torus. We do not see a similar extended velocity structure at
red-shifted velocities. This may be due to the placement of the slits, or may
indicate some internal extinction at that location. The southern tip is
blue-shifted at $-175\rm\,km\,s^{-1}$: similar to the extreme velocity near the
torus, but likely less inclined towards us.

The proper motion of the knots 4 and 5 are around 37 mas/year and 59 mas/year
respectively. For the outer tips of the bipolar lobes, the radial distance to
the central radio source of 35 arcsec indicates a proper motion of 105
mas/year. Our measured proper motion of the southern tip corresponds to
$120\pm25\,\rm mas\,yr^{-1}$.  For a distance of 600\,pc, the respective
tangential velocities are 106 and 169 km/s for the knots, and $340 \rm
\,km\,s^{-1}$ for the tips of the bipolar lobes. Assuming a systemic velocity
$v_{\rm hel} = -50\,\rm km\,s^{-1}$, the space velocity of the tip of the
bipolar lobe becomes $v_{\rm ej} \approx 360 \rm \,km\,s^{-1}$.

The bipolar flow is inclined by 20 degrees with respect to the plane of the
sky, from these numbers.

\subsection{Extended nebula}

The nebula is detected only in ionized lines and in dust. It is possible that
emission lines reveal only part of the nebula, with the remainder not ionized
or too faint. Parameters of the extended nebula must therefore depend on
estimations.

\citet{Shara1985} derive $n_{\rm e} < 10^2\rm \,cm^{-3}$, based on the
[S\,{\sc ii}] line ratio in knot 1.  \citet{Naylor1992} find a higher density
$n_{\rm e} \approx 10^3\rm \,cm^{-3}$, based on the [S\,{\sc ii}] line ratio
of a faint knot (possibly knot 4 or knot 6).  If we assume that the system of
knots 1,2,3,4,6 represents a partially ionized torus with density $n = 10^2\rm
\,cm^{-3}$, diameter of 10 arcsec, thickness and height of 5 arcsec, the total
mass becomes $M\sim 2 \times 10^{-5}\,\rm M_\odot$. If, on the other hand, we
assume that the total extent is given by the bipolar nebula (35 arcsec
radius), and assume a typical density of $n = 10^2\rm \,cm^{-3}$, the total
mass becomes $M\sim 2 \times 10^{-2}\,\rm M_\odot$. The dust mass derived by
\citet{Evans2002a} gives $M \sim 5\times 10^{-2}\,\rm M_\odot$ for a typical
dust-to-gas ratio of 100 (assuming hydrogen-rich gas). Thus, a model of a
cloud of radius 30 arcsec and average density $n = 10^2\rm \,cm^{-3}$ is
consistent with observational constraints.

\citet{Shara1985} suggests that the extended nebula was ionized during the
1670 eruption and is now recombining. This assumption of a slowly recombining
nebula presents some problems.  N$^+$ recombines much faster than hydrogen, so
the high [N\,{\sc ii}]/H$\alpha$ ratio would require that no significant
recombination has yet occurred even in the denser knots.  For [N\,{\sc ii}],
at a density of 100\,cm$^{-3}$ the recombination time scale is 75\,yr
\citep{Lechner2004}.  Densities of $n_{\rm e} \gsim 10^2\rm \,cm^{-3}$ require
continuous ionization.

The line ratios listed by \citet{Shara1985} are in good qualitative agreement
with the radiative shock models of \citet{McKee1980}, although H$\alpha$ is
somewhat fainter than seen in those models. In particular, the [S\,{\sc ii}]
6716+6730/H$\alpha\sim 1$, [N\,{\sc i}] 5200/H$\beta \sim 1$ and the strength
of the [N\,{\sc ii}] 6584\AA\ line are indicative of shock velocities $v \sim
100\,\rm km\,s^{-1}$. (\citet{Shara1985} do not list or identify the [N\,{\sc
i}] 5200\AA\ line, but it is apparent in their spectrum.) The presence of
[O\,{\sc iii}], if shock-ionized, indicates shock velocities above $100\rm
\,km\,s^{-1}$.

\section{The compact radio source}

\subsection{Emission mechanism}

The fact that we have a compact but resolved source suggests thermal
free--free emission for the origin of the radio emission. This is supported by
the observed brightness temperature.  At 5\,GHz, the observed diameter (0.11
arcsec FWHM of a deconvolved gaussian) gives a brightness temperature of
$T_{\rm b} = 6000\,$K. At 8\,GHz we find $T_{\rm b} = 2000\,$K. Assuming the
electron temperature is $T_{\rm e} = 10^4\,K$, the optical depth at 5\,GHz is
$\tau \approx 0.9$.  This predicts at 8\,GHz that $T_{\rm b} \approx 2700\,$K,
in reasonable agreement with observations.  The emission is optically thick at
1.4\,GHz, explaining the non-detection at this frequency \citep{Bode1987}.

The lack of an infrared counterpart presents a problem, however.  Optically
thin free--free emission predicts approximately 0.6\,mJy at 8$\mu$m, assuming
a $\nu^{-0.1}$ index.  The upper limit for a counterpart to the radio source
at 8$\mu$m is around 0.2 mJy.  The only possible non-stellar source is a
possible faint source 2 arcsec south-east of the radio source, which is absent
from the Spitzer images at shorter wavelengths.  The lack of an 8$\mu$m
detection can be explained by very high extinction. However, it also suggests
to explore different emission mechanism.

Small grains can emit appreciable radio emission through their rotation. The
electric dipole emission is described in \citet{Draine1998} and references
therein. Using emissivities per hydrogen atom tabulated in \citet{Draine1998},
we find that our radio flux would require $\sim 3 \times 10^{-3}\,\rm M_\odot$
of gas (assuming normal composition of the gas).  This is a plausible value
for the radio source, and would give densities around $n \sim 10^{7}\,\rm
cm^{-3}$.  However, this emission produces a positive spectral index between 5
and 8\,GHz, while we find a flat flux distribution.  There is also no evidence
for the required very small grains in the infrared spectral energy
distribution.

Synchrotron and cyclotron emission \citep{Linksy1996}, from relativistic
electrons moving in a magnetic field, give brightness temperatures much higher
than we observe. However, if our radio source is found not to be resolved,
this becomes a possibility.

\subsection{Density and mass of the radio core}

If the radio emission is due to free--free emission, we can derive some
parameters of the originating nebula.  Using the observed flux and assuming
$\tau=0.9$ and $T_{\rm e }= 10^4\,$K at 5\,GHz, the emission measure becomes
${\rm EM} = n_{\rm e}^2 l = 8.1 \times 10^7\, \rm pc \,cm^{-6}$ where $n_{\rm
e}$ is the electron density and $l$ the depth of the emitting region.  The
latter is taken as the same as the FWHM of the radio source (assuming
symmetry), or 66\,AU at 600\,pc.

We arrive at a density of $N_{\rm e} = 5 \times 10^5\,\rm cm^{-3}$.  The mass
of the ionized region becomes $M_{\rm i} \approx 4 \times 10^{-7}\,\rm
M_\odot$. This assumes a H-rich chemical composition: H-poor gas requires
more mass per electron.

For a filling factor equal to unity, a relation between the hydrogen line flux
and the optically thin free--free $S_{5{\rm GHz}}$ flux density is given by
\citet{Pottasch1984} assuming hydrogen-rich material.
This gives an expected $F_0({\rm H}\alpha) \approx 1.5
\times 10^{-12}\,\rm erg\,s^{-1}\,cm^{-2}$. The IPHAS H$\alpha$ filter has an
equivalent width of approximately 80\AA\ and a peak transmission of 90 per
cent \citep{Drew2005}: the unreddened magnitude of the central object in the
H$\alpha$ filter should have been $m \sim 12.3$. The (undetected) object is
much weaker. The emission located 0.5 arcsec away from the radio source is
$1.6 \times 10^{-15} \rm erg\,s^{-1}\,cm^{-2}$, three orders
fainter. This requires $A_R \geq 7.5$, corresponding to $A_V \geq 10$ for
a standard extinction law and pure $\rm H \alpha$ emission.

The [N\,{\sc ii}] 6584/6548{\AA} lines are within the IPHAS H$\alpha$ filter
transmission.  However, these lines have a critical density of $n_{\rm cr} = 9
\times 10^4\,\rm cm^{-3}$. We derive an electron density of $n_{\rm e} = 5
\times 10^5\,\rm cm^{-3}$, suggesting that these lines do not contribute
in the central region.

The lack of a clear H$\alpha$ counterpart to the radio source can also be
interpreted as evidence that the core is H-poor.  Hydrogen-poor gas tends to
have lower electron temperatures, as the cooling per nucleus is more
efficient. We ran some {\it Cloudy} models to obtain indicative values. The
models are similar to those described in \citet{Hajduk2005}, but use a PG1159
stellar model \citep{Rauch03}, with luminosity $L= 0.5\,\rm L_\odot$, a
density $n_e=5$--$7\times 10^5\,\rm cm^{-3}$ (depending on stellar temperature
$T_{\rm eff}$ ) and an inner radius of 13\,AU. For $T_{\rm eff}= 60$, 80, 100,
120 kK, we find $T_{\rm e}=8100$, 8400, 8700, 9200K, respectively.  These are
lower than corresponding values of H-rich gas, but not dramatically so.  The
physical cause of the relatively high temperatures is the suppression of the
forbidden lines at these high densities, which reduces the cooling rate.  The
model temperatures are within the range derived from the radio free-free
spectrum. But in the absence of measured line ratios, the models are
exploratory only.

\subsection{Luminosity}

The recombination timescale of the radio core is less than a year.  The
central region is therefore continuously ionized: although no central star is
visible, its continuing existence can be inferred.  Recombining gas can also
be ruled out because it would show a very low electron temperature $T_e\sim
10^3\,\rm K$, much lower than the observed brightness temperature.

The luminosity of the ionizing star can be estimated assuming that all photons
above the Lyman limit ionize hydrogen, and that the object emits as a black
body. The number of ionizing photons reaches a maximum at a temperature of
about $T_\ast = 7 \times 10^4\,$K. For this temperature, the required black
body luminosity is $L_\ast = 0.5\,\rm L_\odot$. For other temperatures within
the range 40--100\,kK, the luminosity is within a factor of 1.5 of this value.
Stellar atmosphere models require a little higher luminosity as a black body
emits more ionizing photons than a realistic atmosphere of the same
temperature (e.g., at $T=50\,\rm kK$, a Kurucz model gives 10 \%\ fewer
H-ionizing photons than does a black body).  Geometrical considerations also
indicate that this luminosity is a lower limit, and the real luminosity could
be a few times higher. 

Such a luminosity, combined with a high stellar temperature, gives an
unreddened magnitude of order $V_0 \gsim 17$. No star is detected near
this magnitude.  The ROSAT all sky survey did not detect any source at the
position of CK Vul. \citet{Sevelli2004} reports extreme faintness of CK Vul in
the SWP IUE spectra; the aperture of the instrument included the position of
the radio source.  The lack of detection may be due to a higher stellar
temperature together with high obscuration.

For comparison, at the adopted distance, the peak absolute magnitude of the 
AD1670 eruption was around $M({\rm V})_0 \approx -9$ to $-8$.  The current 
luminosity implied by the radio source is a factor of $10^5$ fainter.

\subsection{Dust}

The infrared spectrum of CK Vul \citep{Evans2002a} shows two components, with
dust temperatures around 550\,K and 25\,K.  Their two-component dust model
yields approximate dust masses of $1.5\times 10^{-9} $ and $5\times
10^{-4}\,\rm M_\odot$.  The mass of the first component agrees well with the
derived ionized mass, assuming normal dust-to-gas ratios.  The model was based
on IRAS flux densities: the large IRAS apertures include both the
core and the extended nebula, and the second component may be identified
with the extended nebula.

 The new high-resolution Glimpse limit is a factor of 100 below the IRAS
12$\mu$m flux.  We ran dust models with parameters $r_{\rm i}=2 \times
10^{14}\,\rm cm$, $r_{\rm o}= 10^{15}\,\rm cm$, density at inner edge $5
\times 10^5\, \rm cm^{-3}$, density distribution $n \propto r^{-1}$, and a
star with $T= 6 \times 10^4\,\rm K$ and $L=0.5 \,\rm L_\odot$. The (assumed
H-rich) shell has a mass consistent with the ionized mass. For silicate dust,
this yields predicted flux densities of $F_8 = 0.1\,\rm mJy$ and $F_{12}=
0.6\,\rm mJy$, well within observational limits. Carbon-rich dust (as assumed
by \citet{Evans2002a} for their hotter dust component) would be several times
brighter at 8$\mu$m.  These exploratory models differ from \citet{Evans2002a}
mainly in that they give lower dust temperatures (200--300\,K).

The peculiar light curve with several dips may be due to epochs of dust
formation.  In support of this, Hevelius comments that during the second peak
(April 1671), the source appeared reddish \citep{Shara1985}. The brightness of
the initial eruption suggest that it did not suffer the line-of-sight
extinction of the current radio source, and that the dust formed afterwards.

\section{The nature of CK Vul}

\subsection{Observational constraints}

The current luminosity and high temperature places the central object high on
the white dwarf cooling track, e.g., a 1\,L$_\odot$ white dwarf has $T_{\rm
eff} \approx 65\,\rm kK$ and a cooling age of $t_{\rm c} \sim 10^6$\,yr
\citep{Bloecker1995}. 

The compactness of the radio source implies it is not expanding, as the
expansion velocity would be less than 1\,km\,s$^{-1}$, or below the escape
velocity.  We can therefore assume that the radio nebula is part of a stable
structure, such as a rotating disk, and could predate the eruption.
The bipolar lobes are similar to those found in some young planetary nebulae
and post-AGB stars. Such lobes can form where a spherical fast wind meets an
existing equatorial obstruction \citep{Icke1989}.

 An area of low dust emission surrounding CK Vul is seen in IRAS images
\citep{Evans2002a}.  This resembles cases where the ISM has been swept up by a
stellar wind \citep{Zijlstra2002a} and suggests CK Vul has gone through phases
of high mass loss over the past $\sim 10^{5}$--$10^{6}$\,yr.

Together, these constraints favour an evolved object, and the morphology
favours a binary object. Two possible interpretations are a nova, and a
thermal pulse.


One should consider whether a supernova classification can truly be ruled out.
The [S\,{\sc ii}]/H$\alpha\sim 1$ ratio is diagnostic of supernova remnants.
Also, the current luminosity is a typical value for a pulsar, and the radio
spectrum is similar to pulsar wind nebulae \citep[e.g.][]{Temim2006}.  The
object H\,2-12, a knot inside the Kepler supernova remant, shows similar line
ratios and velocities to CK Vul \citep{Riesgo2005}.  However, there is no
X-ray source at this position, nor is there significant radio emission from
the extended nebula. The bipolar symmetry of CK Vul differs from the structures
seen in supernova remnants. We conclude that a supernova classification is
unlikely.

A 'gentle' supernova \citep{Tout2006} can be considered. An accreting helium
white dwarf may ignite its helium at masses far below the Chandrasekhar mass
\citep{Woosley1994}. The lack of observational evidence for sub-luminous
supernovae suggests that the ignition under such conditions may be
non-explosive. However, such objects may also have been classified
wrongly. The models of \citet{Woosley1994} predict short-lived maxima, however.

The observations do not clearly support the possibility of a stellar merger
\citep{Kato2003}: a merger would be able to explain the large drop in
luminosity since the eruption, but would lead to a bloated, cool star.  The
radio source requires an ionizing star.

\subsection{A light nova}

Classical novae are dominated by massive white dwarfs: these show frequent
eruptions, but consequently eject relatively little mass.  A lower-mass white
dwarf with a low accretion rate can eject the most massive shell.  However,
such eruptions are rare as a consequence of the much longer accretion time
scales. The current luminosity of CK Vul corresponds to an accretion rate
of $\dot M_{\rm acc} \sim 10^{-10} \,\rm M_\odot\,yr^{-1}$.  
The peak temperature, for a thermal-nuclear runaway on a low-mass white
dwarf, may not even reach $10^8$\,K limiting the nucleosynthesis wich occurs
\citep{Starrfield2007}.

A large set of models covering a range of parameter space is presented by
\citet{Prialnik1995}. Some of these approach the constraints set by CK
Vul. For instance, their model with $M_{\rm WD}=0.65\,\rm M_\odot$, $T_{\rm
WD}=3 \times 10^7\,\rm K$ and $\dot M = 10^{-10}\,\rm M_\odot\, yr^{-1}$
yields an ejection velocity of 200\,km\,s$^{-1}$, eruption amplitude of
16\,mag, mass-loss time scale of 480 days (during which time the nova remains
bright), and a very high peak luminosity of $L=8 \times 10^4\,\rm L_\odot$
($M_{\rm bol}=-7.5$). The recurrence time scale of this model is
$10^6$\,yr. 

This particular model gives an ejecta mass of $10^{-4}\,\rm M_\odot$. The
highest prediced ejecta mass for a light nova is only $7 \times 10^{-4}\,\rm
M_\odot$ \citep{Yaron2005}.  On the one hand, this is three orders of
magnitude higher than found in near-Chandrasekhar-mass novae. On the other
hand, it is one to two orders of magnitude below the derived nebular mass of
CK Vul (and below the mass in the dust alone). It is however possible that
part of this is swept-up material which predates the eruption. 

\subsection{A thermal pulse}

Helium ignition on the white dwarf cooling track has been observed in V605 Aql
\citep{Clayton2006} and V4334 Sgr \citep[Sakurai's Object:][]{Asplund1999,
Hajduk2005, Evans2006}. Their light curves show similarities to CK Vul: a long
lived maximum at $M\sim -5$, with a brief phase of very high, dusty mass loss.
Three further objects are known to have experienced a VLTP event in their
evolution, as shown by the presence of hydrogen-poor central regions within a
planetary nebula \citep{Zijlstra2002}. Outflow Velocities are also similar,
with observed values are $\sim 250  \,km\,s^{-1}$ for V605 Aql 
\citep{Pollacco1992} and  V4334 Sgr \citep{Kerber2002}.

If CK Vul is a VLTP, it would fall on an evolutionary sequence V4334 Sgr
(Sakurai's Object: 10\,yr), V605 Aql (90\,yr), IRAS\,15154$-$5258 ($\sim
10^3$\,yr) and the two more evolved objects A30/A78 \citep{Zijlstra2002}. The
first two are still hidden inside a dense cocoon of dusty ejecta, but like CK
Vul the central object is seen in radio emission.  There are some
morphological resemblances with CK Vul, with the ejecta in these objects
showing equatorial torii and polar knots. The origin of the morphologies are
not known.

There are however also notable differences. The radio source in CK Vul is much
more compact than expected: in V605 Aql, which is at a larger distance, the
0.3-arcsec dense shell is expanding and there is no evidence for a
non-expanding component.  The ejecta in both VLTP objects are almost
completely H-poor. In contrast, most knots in CK Vul show significant
H$\alpha$ components, even if fainter than expected.  (We do not detect
H$\alpha$ emission in knot 4, which is likely H-poor. It is possible that the
other knots have swept up ambient hydrogen.)  CK Vul has a much lower current
luminosity than the other five objects. Finally, fossil planetary nebulae are
seen in all five objects, but CK Vul only shows a surrounding hole in the ISM
as evidence for past mass loss: this suggests a much longer post-AGB time
scale before its eruption.

VLTP events may show a characteristic double loop in the HR diagram, over a
time scale of a few hundred years \citep{Hajduk2005}.  Direct observational
evidence for a second loop is still lacking and the VLTP scenario is not well
understood \citep{Miller2006}.  Observations during the 18$^{\rm th}$ and
19$^{\rm th}$ century \citep{Shara1985} provide no indications for a second,
decadal phase of high luminosity and low temperatures for CK Vul, as predicted
by double-loop models, but the extinction may have hidden such an event.
However, the low luminosity we derive is a severe constraint for such models.

\subsubsection{Accretion-induced pulse}

The morphology shows similarities to evolved
binary systems, such as symbiotics. This together with the compact radio
source suggests variation on the standard VLTP scenario can be considered,
involving accretion.

 A group of binary post-AGB stars, including RV Tau stars, show compact
circumstellar disks, with typical sizes of 10-100 AU \citep{vanWinckel2003,
deRuyter2006}. The disks form when a close binary ($P\sim 1\,\rm yr$) captures
matter lost by the central star into a circumbinary disk.  The radio source in
CK Vul has the correct size to be the remnant of such a disk.  CK Vul shows
similarities in particular with OH231.8+4.2, which shows bipolar lobes, an
inner disk with diameter of 50\,AU as well as a 1000\,AU dense torus
\citep{Matsuura2006}.

 In such a system, the post-AGB evolution may be affected by accretion from
the disk on the post-AGB star \citep{Zijlstra2001}. Such accretion could
trigger a thermal pulse later during the post-AGB evolution than wold
otherwise occur. The circumstellar environment can explain the bipolar
outflows.  This possibility should be explored further for CK Vul.

\section{Conclusions}

A compact radio source allows us for the first time to identify
the central object of the old nova CK Vul. The radio emission is modelled as
thermal free-free emission, from a compact nebula of diameter $\sim
60\,$AU. The density of the nebula is $n_{\rm e} = 5 \times 10^5\,\rm
cm^{-3}$.  A high extinction through the circumstellar shell is indicated by
the lack of H$\alpha$ detection at the position of the radio source. The
ionization requires a luminosity of the ionizing object of order
1\,L$_\odot$.

Deep H$\alpha$ images reveal a large bipolar outflow, with a diameter of 70
arcsec. The nebulosity is brightest near the position of the radio source, and
shows various components. Comparison with previous images shows significant
expansion of the nebula. Extrapolating back shows that the $H\alpha$
structures were ejected in the AD 1670 eruption. The centre of expansion is,
within the uncertainties, identical to the radio source. Echelle spectra and
proper motions give original ejecta velocities of $\sim 360 \rm\, km\,s^{-1}$.

The observed characteristics
differ from classical novae. New models of nova eruption on cool, low-mass
white dwarfs can explain some of the observational constraints.  A
classification as a Very Late Thermal Pulse appears possible, possibly
accretion-induced. The radio source is proposed to be the remnant of a
circumbinary disk, as seen in some binary post-AGB stars. The ejecta
morphology are suggestive of binary interaction. The radio source provides
important information on the environment of CK Vul. However, at present CK Vul
still remains an enigma \citep{Evans2002a, Naylor1992, Shara1985}.

\section*{Acknowledgments} We thank Antonio Garcia, Kerttu Viironen 
and Ramarao Tata for carrying out the INT observations.  The VLA is part of
the National Radio Astronomy Observatory, operated by AUI inc on behalf of the
NSF. The INT is part of the ING group of telescopes.  This project was
supported by a NATO grant for collaborative research, by PPARC through a
rolling grant, and by Uniwersytet Mikolaja Kopernika through grant UMK
366-A. PvH acknowledges support from the Belgian Science Policy Office through
grant MO/33/017.  JAL and MGR acknowledge financial support from CONACyT grant
43121 and UNAM-DGAPA grants IN108506-2, IN108406-2, and IN112103. AAZ thanks
the SAAO for hospitality during a sabbatical visit.




\label{lastpage}

\end{document}